\documentclass[conference]{IEEEtran}
\IEEEoverridecommandlockouts
\usepackage{cite}
\usepackage{amsmath,amssymb,amsfonts}
\usepackage{algorithmic}
\usepackage{array}
\usepackage{amssymb}
\usepackage{graphicx}
\usepackage{svg}
\usepackage{textcomp}
\usepackage{tikz}
\usepackage{multirow}
\usepackage{enumitem}

\usepackage{colortbl}
\usepackage{xcolor}
\def\BibTeX{{\rm B\kern-.05em{\sc i\kern-.025em b}\kern-.08em T\kern-.1667em\lower.7ex\hbox{E}\kern-.125emX}}
\newcolumntype{C}[1]{>{\centering\arraybackslash}m{#1}}

\begin{document}
\setlength{\arrayrulewidth}{0.2mm}
\arrayrulecolor{black}
\setlength{\abovedisplayskip}{3pt plus 0pt minus 1pt}
\setlength{\belowdisplayskip}{3pt plus 0pt minus 1pt}
\setlength{\abovecaptionskip}{3pt plus 0pt minus 1pt}
\setlength{\belowcaptionskip}{3pt plus 0pt minus 1pt}
\setlength{\textfloatsep}{3pt plus 0pt minus 1pt}
\setlength{\floatsep}{3pt plus 0pt minus 1pt}
\setlength{\dbltextfloatsep}{3pt plus 0pt minus 1pt}
\setlength{\dblfloatsep}{3pt plus 0pt minus 1pt}
\title{An Event-Based Digital Compute-In-Memory Accelerator with Flexible Operand Resolution\\ and Layer-Wise Weight/Output Stationarity
\vspace*{-0.25cm}}

\author{\IEEEauthorblockN{Nicolas Chauvaux\textsuperscript{1}, Adrian Kneip\textsuperscript{1,2}, Christoph Posch\textsuperscript{3}, Kofi Makinwa\textsuperscript{1}, and Charlotte Frenkel\textsuperscript{1}
\\
\textsuperscript{1} Department of Microelectronics, Delft University of Technology, The Netherlands\\
\textsuperscript{2} ESAT Department, KU Leuven, Belgium  \hspace{0.15cm} \textsuperscript{3} Prophesee, 75012 Paris, France
\vspace*{-0.4cm}
}
}

\maketitle

\begin{abstract}
Compute-in-memory (CIM) accelerators for spiking neural networks (SNNs) are promising solutions to enable $\mu$s-level inference latency and ultra-low energy in edge vision applications. Yet, their current lack of flexibility at both the circuit and system levels prevents their deployment in a wide range of real-life scenarios. In this work, we propose a novel digital CIM macro that supports arbitrary operand resolution and shape within a unified CIM storage for weights and membrane potentials. These circuit-level techniques enable a hybrid weight- and output-stationary dataflow at the system level to maximize operand reuse, thereby minimizing costly on- and off-chip data movements during the SNN execution. Measurement results of a fabricated FlexSpIM prototype in 40-nm CMOS demonstrate a 2$\times$ increase in 1-bit-normalized energy efficiency compared to prior fixed-precision digital CIM-SNNs, while providing resolution reconfiguration with bitwise granularity. Our approach can save up to 90\% energy in large-scale systems, while reaching a state-of-the-art classification \mbox{accuracy of 95.8\% on the IBM DVS gesture dataset.}
\end{abstract}


\vspace{-0.1cm}
\section{Introduction}
\begin{figure}[!htbp]
\centerline{\includegraphics[width=\linewidth]{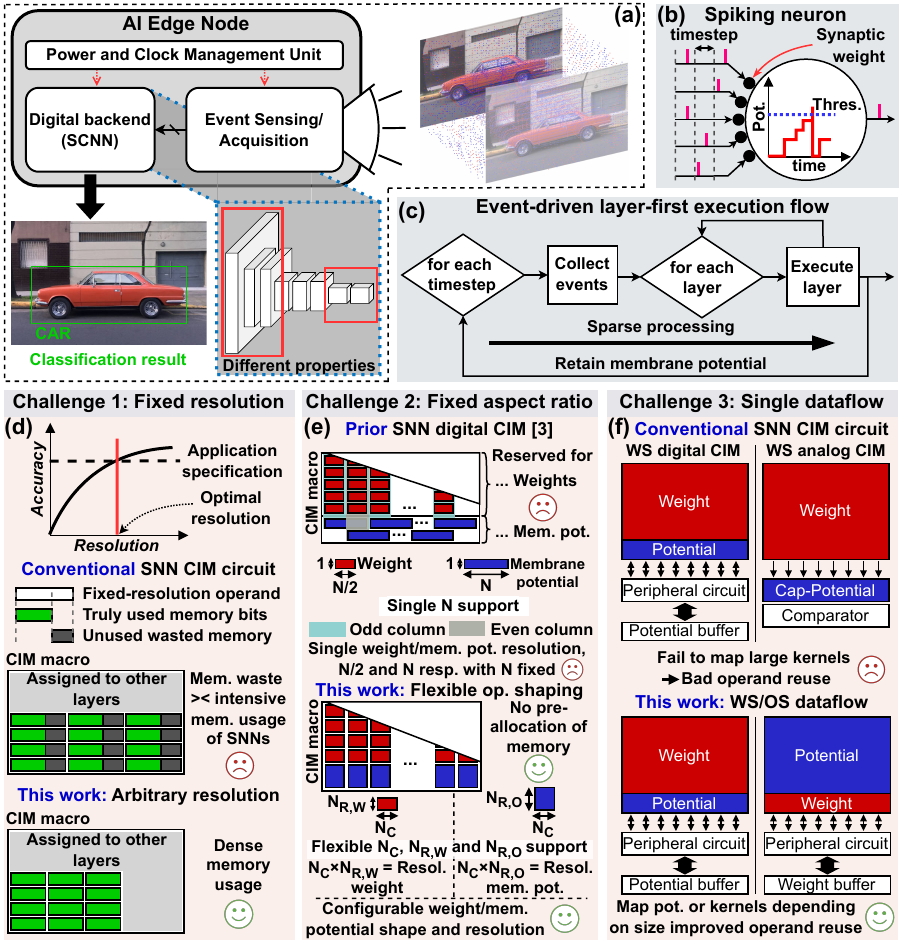}}
\caption{(a)~Event-based edge vision system and workload example. (b)~Integrate-and-fire spiking neuron model. (c)~Adopted execution flow targeting low latency execution. (d-f)~Three core challenges in the state of the art and corresponding innovations in the proposed design.}
\label{fig:challenges}
\end{figure}
Deploying convolutional neural networks (CNNs) to enable vision at the edge has unlocked applications ranging from user recognition to decentralized object detection and classification. Recently, event-based vision has emerged as a promising opportunity for reduced system-level energy and latency \cite{IBM}. Indeed, pixels of event-based cameras fire events independently, generating sparse event streams at a $\mu$s-level temporal resolution (Fig.~\ref{fig:challenges}(a)), which calls for dedicated algorithms \cite{DVSAlgo}. Among them, spiking neural networks (SNNs) exploit bio-inspired neuron models (e.g., integrate-and-fire (IF) in Fig.~\ref{fig:challenges}(b)) to sparsely process information using binary spikes: they rely on an internal state called membrane potential to retain the evolution of information across time. While SNNs are well suited to map per-timestep processing scenarios for low-latency decisions at the edge (Fig.~\ref{fig:challenges}(c)), large-scale models that cannot fit entirely in the on-chip memory suffer from significant data movement overheads, which prevents their efficient deployment on conventional hardware architectures.%\\

\begin{figure}[htbp]
\centerline{\includegraphics[width=\linewidth]{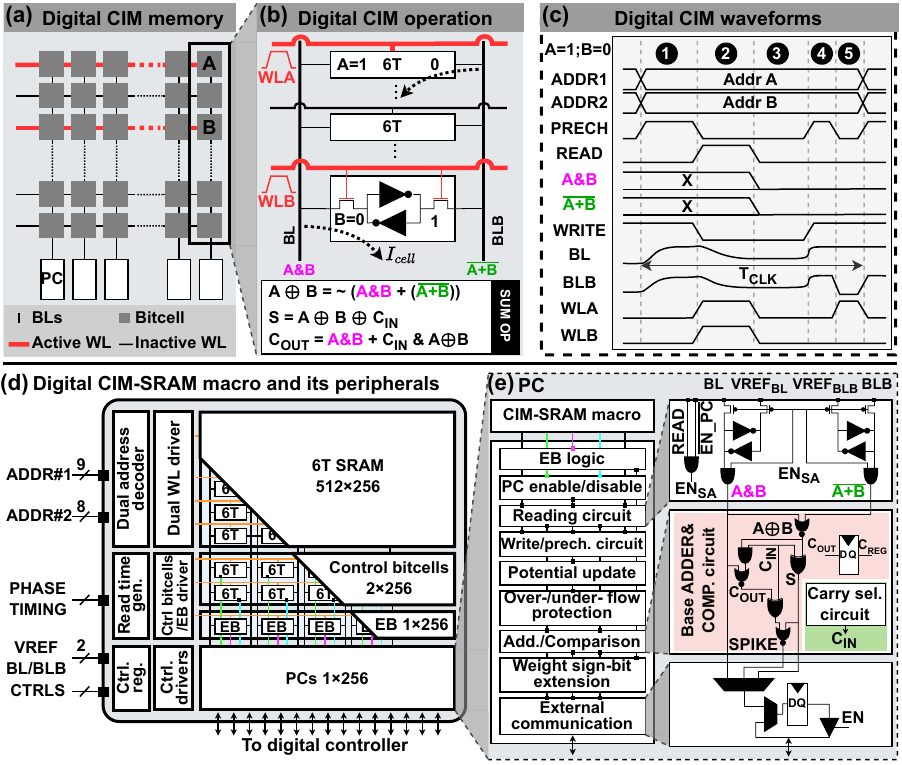}}
\caption{(a)~Digital CIM memory overview: array of bitcells and peripheral circuit (PC) attached to each bitline (BL) for operation handling. Following digital CIM operations, two wordlines (WLs) are simultaneously activated. (b)~Example of a digital CIM operation between two bitcells storing A=1 and B=0. Two boolean operations are obtained and can be used to obtain a 1-bit full adder by following the equations provided. (c)~Example of waveforms and phases to perform the digital CIM operation illustrated in (b). (d)~Architecture of the proposed FlexSpIM digital CIM-SRAM macro. (e)~Decomposition of a PC into modules with their detailed schematics.}
\label{fig:combined_cim_details}
\end{figure}

\indent To that end, compute-in-memory (CIM) hardware for SNNs has recently been proposed \cite{IMPULSE, ISSCC24, NeuroCIM, SpikeCIM, DRAM_CIM, clock_free_cim, ds_cim, ISCAS22, SNNIM, SRIF}, but their limited reconfigurability makes them fundamentally unsuited to the large diversity of CNN layer specifications (Fig. \ref{fig:challenges}(a)). This either implies sub-optimal workload mapping to CIM hardware, leading to latency and energy penalties, or increased time-to-market due to the need for application-specific CIM hardware.
In this work, we propose FlexSpIM, a digital CIM-based accelerator for SNN inference with high flexibility at the circuit and system levels. It solves three core challenges compared to prior CIM-based SNNs:
\vspace*{-1mm}
\begin{enumerate}[leftmargin=*]
    \item while previous works only support a fixed resolution or a few pre-defined options, FlexSpIM supports a fully reconfigurable resolution for weights and membrane potentials, thereby expanding the exploration of the trade-off landscape between accuracy, energy efficiency, and memory footprint for SNN workloads~(Fig.~\ref{fig:challenges}(d));
    \item while operand mapping is usually restricted to either fully bit-serial row-wise or fully bit-parallel column-wise, FlexSpIM allows for reconfigurable operand shapes to support different, non-proportional resolution values for weights and membrane potentials, which were otherwise constrained to fixed ratios~\cite{IMPULSE}~(Fig.~\ref{fig:challenges}(e));
    \item while maximizing operand stationarity is key to reduce external memory accesses, prior CIM-based SNNs only support weight stationarity, and thus are ill-suited for layers that are bottlenecked by membrane-potential data movement (e.g., first layers of ResNet \cite{ResNet}). FlexSpIM introduces a unified weight/membrane potential memory that allows for a hybrid dataflow, making the best out of both weight stationarity (WS) and output (i.e., membrane potential) stationarity (OS) on a per-layer basis. This minimizes operand replacement in the CIM macro for the selected workload, thereby directly alleviating the data movement efficiency bottleneck of previous work~(Fig.~\ref{fig:challenges}(f)).
\end{enumerate}

\begin{figure}[!htbp]
\centerline{\includegraphics[width=\linewidth]{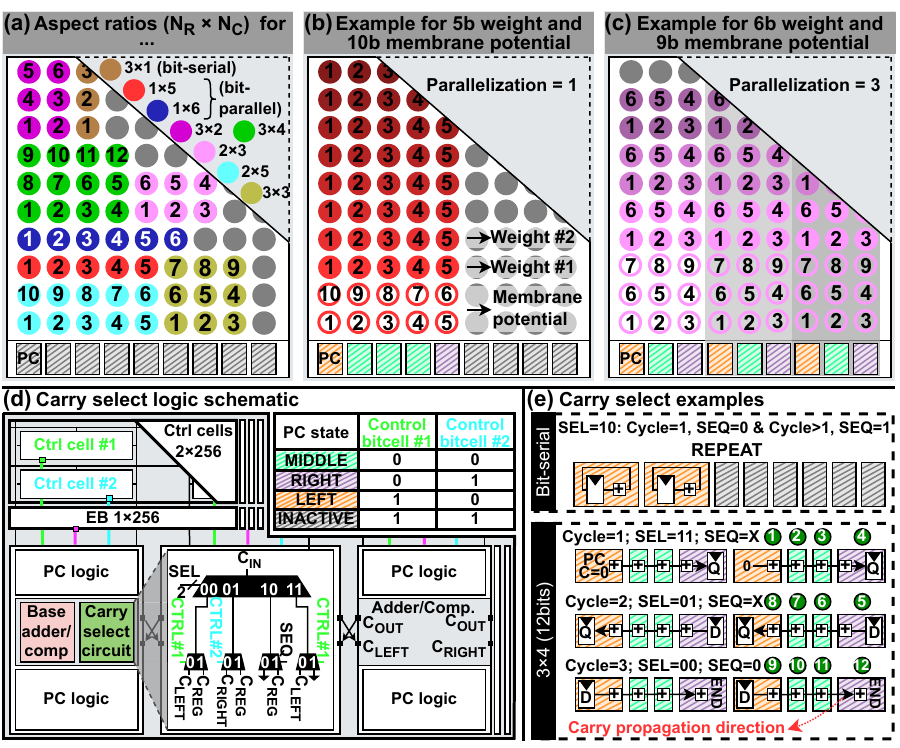}}
\caption{(a)~Arbitrary resolution and operand shaping principle. (b)~Example of operand shaping for 5-bit weight and 10-bit membrane potential with a selected parallelization of one neuron. (c)~Example of operand shaping for 6-bit weight and 9-bit membrane potential with a selected parallelization of three neurons. (d)~Carry-selection logic and PC state configuration modes. (e) PC configurations for bit-serial and 4$\times$3 operand shaping.}
\label{fig:combined_flexible_resolution}
\end{figure}

\section{Proposed Reconfigurable Digital CIM Macro}
SRAM-based CIM is a computing paradigm where operations are carried out directly inside the SRAM, harnessing a low-cost reuse of the data stored in the array. In the case of boolean digital CIM-SRAMs \cite{NeuralCache}, multiple bitwise operations can be performed in parallel on the different vertical bitlines (BLs) by enabling two shared horizontal wordlines (WLs) at a time (Figs.~\ref{fig:combined_cim_details}(a) and (b)). Applied to SNNs, these architectures can execute the XNOR-and-accumulate operation of IF neurons (Fig. \ref{fig:challenges}(b)) by sequentially accessing corresponding bits of a weight A and a membrane potential B, and updating the potential accordingly \cite{IMPULSE}. In this work, we split this operation in five phases (Fig. \ref{fig:combined_cim_details}(c)): 1)~precharge of BL/complementary BL (BLB) to VDD, 2)~AND/NOR operation between the stored A and B operand values on BL/BLB by activating the corresponding WLs, 3)~1-bit sum and carry-out generation in the peripheral circuit (PC) yielding the updated membrane potential, 4)~half-select-prevention BL/BLB precharge, and 5)~write-back of the new membrane potential bit into A. These steps are repeated until all bits of the membrane potential and weight have been processed. Then, a comparison between the resulting membrane potential and a threshold conditionally generates an output spike, transferred to the next layer.

Reconfigurability in the FlexSpIM macro (Fig.~\ref{fig:combined_cim_details}(d)) is achieved by combining a dense 6T SRAM array storing both the weights and membrane potentials with a modular PC per column (Fig.~\ref{fig:combined_cim_details}(e)). Two additional control bitcells define the per-PC state, while emulation bits (EBs) allow for sign-bit extension and write-free CIM operation during data broadcasting in the macro. Each pitch-matched PC consists of a dual sense amplifier (SA) for the individual readout of BL/BLB, a 1-bit full-adder adapted from \cite{NeuralCache} with a comparison and a carry-in selection circuits, and logic for I/O communication.

\subsection{Arbitrary Operand Resolution and Shape}
FlexSpIM leverages arbitrary operand resolutions (i.e., 1-to-512$\times$256-bit with bitwise granularity), which can be selected on a per-layer basis for both weights and membrane potentials (Fig.~\ref{fig:combined_flexible_resolution}(a)). This degree of flexibility overcomes the limited set of resolutions supported in previous works \cite{IMPULSE, ISSCC24, NeuroCIM, SpikeCIM, DRAM_CIM, clock_free_cim, ds_cim, ISCAS22, SNNIM, SRIF}, thereby preventing any waste of storage space. To support weight and membrane potential operands that may take different and non-proportional resolutions, the carry-select circuit (Fig.~\ref{fig:combined_cim_details}(e)) allows FlexSpIM to map operand bits using any shape in the unified 6T SRAM array (Figs.~\ref{fig:combined_flexible_resolution}(b) and (c)). The number of columns occupied by each multi-bit operand is defined using the 2-bit control bitcells, which allow chaining multiple 1-bit adders of neighboring PCs for multi-bit computation by changing the carry-in origin (Fig.~\ref{fig:combined_flexible_resolution}(d)). The multi-bit CIM operation then occurs in parallel over the columns, and sequentially from the LSB row to the MSB row (Fig.~\ref{fig:combined_flexible_resolution}(e)). For multi-cycle operations, a ping-pong left/right sum direction is used between subsequent cycles to keep inter-PC data movement bounded to their direct neighbors, ensuring design scalability to any macro dimensions.

\begin{figure}[t]
\centerline{\includegraphics[width=\linewidth]{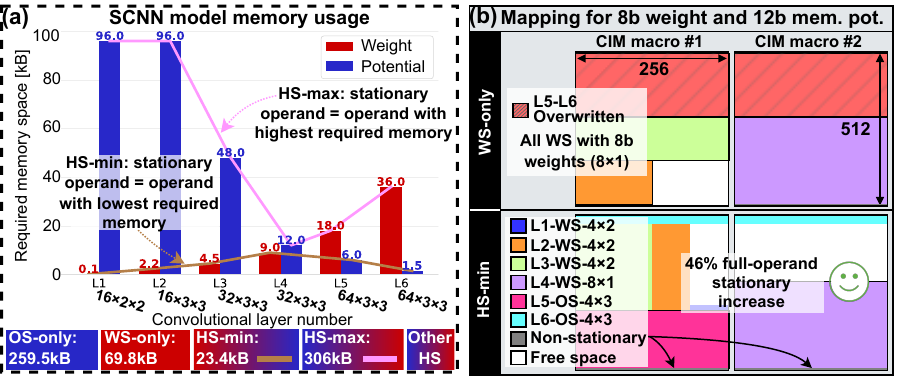}}
\caption{(a)~Layer-level memory requirements for weights and membrane potentials of a spiking CNN composed of six convolutional layers (i.e., L1 to L6) and three FC layers (not shown), with two different HS dataflow situations highlighted by the brown and pink lines. (b)~Mapping of the model on two CIM macros for WS-only and HS-min dataflows.}
\label{fig:model_size_and_two_macros_mapping}
\end{figure}

\subsection{Hybrid-Stationary Dataflow}
At the system level, the unified memory of FlexSpIM allows to support hybrid stationarity (HS) by choosing between WS and OS on a per-layer basis, contrary to previous WS-only CIM designs for SNNs \cite{IMPULSE, ISSCC24, NeuroCIM, SpikeCIM, ds_cim, ISCAS22, SNNIM, SRIF}. To illustrate this point, we consider a typical spiking CNN workload (SCNN), made of six convolution layers (defined in Fig.~\ref{fig:model_size_and_two_macros_mapping}(a)) followed by three fully-connected (FC) ones (not shown). Exploiting the known memory requirements of both weight and membrane potential operands in each of the SCNN's layers, the HS flow selects each layer’s dataflow type in order to maximize the overall utilization of the CIM storage space (Figs.~\ref{fig:model_size_and_two_macros_mapping}(a) and (b)), thereby increasing the overall operand stationarity across the multi-timestep execution of a model. This principle can be extended to multiple macros, where a full HS scenario requires at least two macros to ensure the full stationarity of at least one of the operands of every layer when targeting our SCNN workload. Fig.~\ref{fig:model_size_and_two_macros_mapping}(a) depicts two HS dataflows in which the stationary operand is either the one requiring the least or the most memory, respectively referred to as HS-min and HS-max. Compared to the conventional WS-only dataflow, HS-min increases the amount of stationary operands by $46\%$ with an optimal layer mapping across both macros (Fig. \ref{fig:model_size_and_two_macros_mapping}(b)). Further efficiency gains can be unlocked by scaling up the number of macros, where additional CIM storage avoids frequent external memory accesses by ensuring the stationarity of the operands with the largest memory footprint (Fig. \ref{fig:model_size_and_two_macros_mapping}(a)).

\section{Implementation and measurement results}
\begin{figure}[!t]
\centerline{\includegraphics[width=\linewidth]{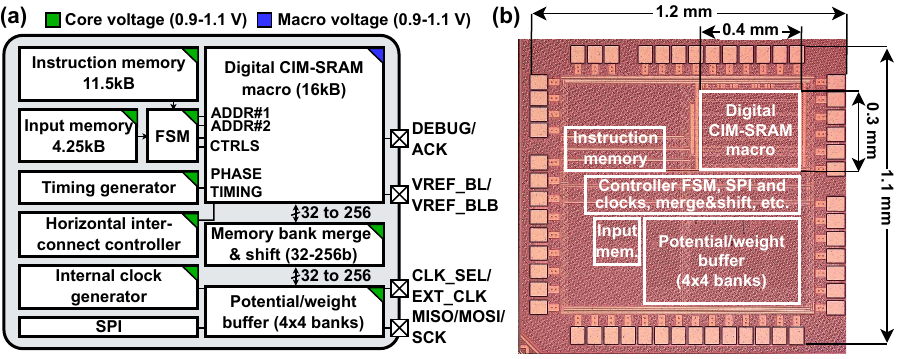}}
\caption{(a) Overall FlexSpIM system architecture and (b) chip microphotograph in bulk 40-nm CMOS.\vspace*{2mm}}\label{fig:overall_architecture_and_microphotograph}
\end{figure}
\begin{figure}[!t]
\centerline{\includegraphics[width=\linewidth]{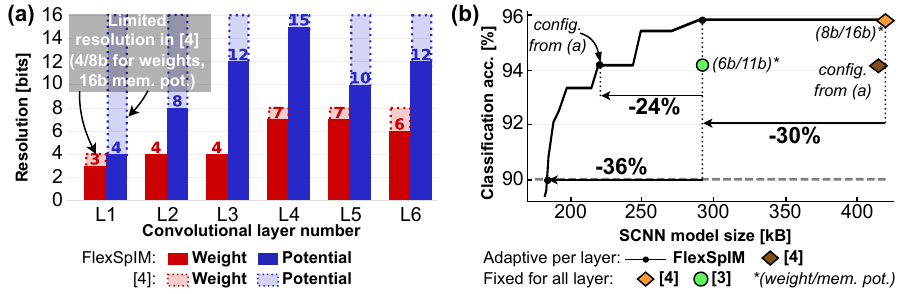}}
\caption{(a)~Weights and membrane potentials resolution of the target spiking CNN to achieve a $94.2\%$ accuracy on the IBM DVS gesture dataset with lowest model size supported by FlexSpIM (unconstrained) and \cite{ISSCC24} (constrained due to lack of flexibility). (b)~IBM DVS gesture sensitivity to resolution for the same spiking CNN and impact on the model size (excluding FC layers).}
\label{fig:resolution_model}
\end{figure}

The proposed CIM macro is integrated within the complete accelerator shown in Fig. \ref{fig:overall_architecture_and_microphotograph}(a). Beyond the proposed 16kB CIM-SRAM macro, it is composed of a 4.25kB memory for per-timestep input spike buffering, and 4$\times$4 banks of 2kB SRAMs buffering the SNN weights (resp.~membrane potentials) in OS (resp.~WS) mode. A 32-to-256-bit bandwidth-adaptive merge-and-shift unit ensures correctly aligned data transfers for arbitrary CIM configurations. The microphotograph of the chip in bulk 40-nm CMOS is shown in Fig.~\ref{fig:overall_architecture_and_microphotograph}(b).

\begin{figure*}[htbp]
\centerline{\includegraphics[width=\linewidth]{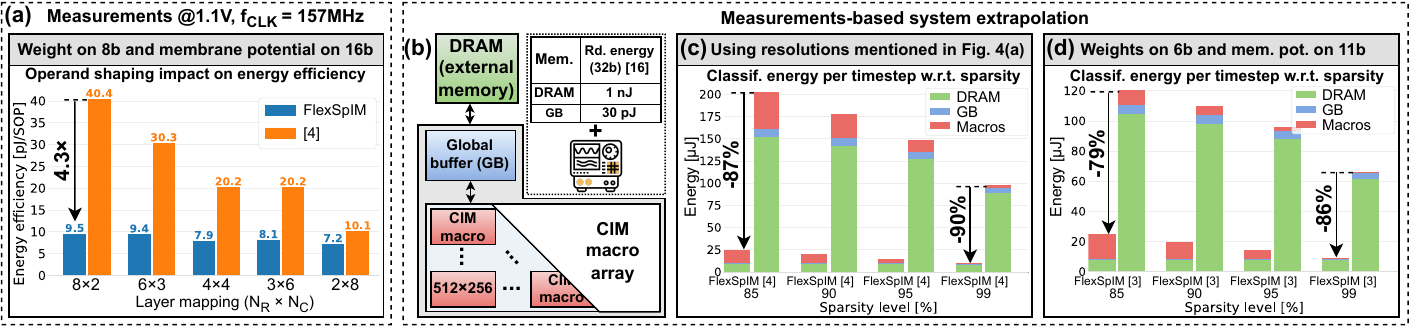}}
\caption{(a) Measurement results of shape-dependent energy efficiency. (b) System architecture used for the many-macro extrapolation. (c-d) System-level many-macro extrapolation comparing FlexSpIM to \cite{ISSCC24} and \cite{IMPULSE}, respectively.}
\label{fig:results}
\end{figure*}

\begin{table*}[!htbp]
\caption{Comparison to the state of the art of SNN accelerators.}
\vspace*{-0.5cm}
\begin{center}
\begin{footnotesize}
\begin{tabular}{|C{0.21\linewidth}!{\color{blue}\vrule width 0.4mm}C{0.105\linewidth}!{\color{blue}\vrule width 0.4mm}C{0.115\linewidth}|C{0.105\linewidth}|C{0.105\linewidth}|C{0.105\linewidth}|C{0.105\linewidth}|}
\arrayrulecolor{black}
\cline{1-1}\cline{3-7}
\arrayrulecolor{blue}\cline{2-2}
\arrayrulecolor{black}

\textbf{} & \textbf{This work} & SSC-L'21 \cite{IMPULSE} & ISSCC'24 \cite{ISSCC24} & JSSC'23 \cite{NeuroCIM} & A-SSCC'22 \cite{SpikeCIM}& ISSCC'22 \cite{ReckOn}\\
\hline
Technology & 40nm & 65nm & 22nm & 28nm & 65nm & 28nm\\
Implementation & \textbf{Digital (CIM)} & \textbf{Digital (CIM)} & Analog CIM & Analog CIM & Analog CIM & Digital\\
Core area (mm$^2$) & 1.37 & 0.089 $^{\textbf{a}}$ & 2.28 & 2.9 & 0.25 $^{\textbf{a}}$ & 0.45\\
Macro memory capacity (kB) & 16 & 1.37 & 4 & 20 & 4 & N/A\\
Bitcell type & 6T & 10T & 6T & 8T & 2$\times$6T+6T & N/A\\
\hline
Spiking network type & CNN & Modified LeNet5 & Residual CNN & ResNet-12 & CNN & RNN\\
Representative dataset & DVS gesture $^{\textbf{b}}$ & MNIST/IMDB & DVS gesture $^{\textbf{b}}$ & CIFAR-10 & MNIST/CIFAR10 & DVS gesture $^{\textbf{b}}$\\
Accuracy on DVS gesture & $95.8\%$ & N/A & $94\%$ & N/A & N/A & $87.3\%$\\
\hline
Multi-aspect ratio support & \textbf{\checkmark} & $\times$ & $\times$ & $\times$ & $\times$ & $\times$\\
HS support & \textbf{\checkmark} & $\times$ & $\times$ & $\times$ & $\times$ & $\times$\\
Mem. pot. resolution & \textbf{Any} & 11b & 16b & 8b & Analog & 16b\\
Weight resolution & \textbf{Any} & 6b & 4/8b & 1/4/8b & 1.5b & 8b\\
Weight/Mem. pot. location & \textbf{Not fixed} & Fixed & Fixed & Fixed & Fixed & Fixed\\
\hline
Supply range (V) & 0.9 -- 1.1 & 0.7 -- 1.2 & 0.55 -- 0.9 & 1.1 & N/A & 0.5 -- 0.8\\
Frequency (MHz) & 75.5 -- 157 & 66.7 -- 500 & 51 -- 280 & 200 & N/A & 13 -- 115\\
Peak throughput (GSOPS)$^{\textbf{*}}$ & 1.2 -- 2.5 $^{\textbf{c}}$ & 0.07 -- 0.5 $^{\textbf{d}}$ & N/A & N/A & 163.8 $^{\textbf{f}}$ & 0.013 -- 0.115 $^{\textbf{c}}$\\
1b-norm. throughput (GSOPS)$^{\ddag}$ & 154 -- 320 & 4.62 -- 33 & N/A & N/A & N/A & 1.67 -- 14.7\\
Power (mW) & 6.8 -- 17.9 $^{\textbf{a}}$ & 0.1 -- 0.9 $^{\textbf{a}}$ & 0.524 -- 6.4 & 15.84 $^{\textbf{a}}$ & 0.56 $^{\textbf{a}}$ & 0.077 -- N/A\\
Efficiency (pJ/SOP)$^{\textbf{*}}$ & 5.7 -- 7.2 $^{\textbf{c}}$ & 1.09 -- 1.74 $^{\textbf{d}}$ & 3.78 -- 10.01 $^{\textbf{c}}$ & 0.0016 $^{\textbf{e}}$ & 3.45$\times$10$^{-3~ \textbf{(f)}}$ & 5.3 -- 12.8 $^{\textbf{c}}$\\
1b-norm. eff. (fJ/SOP)$^{\dag}$ & 44.5 -- 56.3 & 16.5 -- 26.4 & 29.5 -- 78.2 & 0.025 & N/A & 41.4 -- 100\\
\arrayrulecolor{black}
\cline{1-1}\cline{3-7}
\arrayrulecolor{blue}\cline{2-2}
\arrayrulecolor{black}
\multicolumn{7}{l}{$^{\mathrm{\textbf{a}}}$CIM macro only \quad
$^{\mathrm{\textbf{b}}}$ 10 classes \quad
$^{\mathrm{\textbf{c}}}$ 8-bit weight and 16-bit mem. pot. \quad
$^{\mathrm{\textbf{d}}}$ 6-bit weight and 11-bit mem. pot. \quad
$^\mathrm{\textbf{e}}$ 8-bit weight and 8-bit mem. pot.}\\
\multicolumn{7}{l}{$^{\mathrm{\textbf{f}}}$1.5-bit weight and N/A for mem. pot. \quad
$^{\mathrm{\ddag}}$GSOPS $\times$ weight-bit $\times$ pot.-bit \quad
$^{\mathrm{\dag}}$fJ/SOP/(weight-bit $\times$ pot.-bit) \quad
$^{\mathrm{*}}$1 SOP = 1 addition + mem. pot. update}\\
\end{tabular}
\end{footnotesize}
\label{comparison_table}
\end{center}
\end{table*}
\subsection{Circuit Level}
The unrestricted support of the neural network quantization landscape allows for workload-dependent accuracy/footprint optimization. On the IBM DVS gesture dataset \cite{IBM} and our six-layer SCNN model, we demonstrate that fine-tuning the resolution of the operands allows for a 30$\%$ memory footprint reduction over \cite{ISSCC24} while maintaining a state-of-the-art accuracy of $95.8\%$ for gesture classification (Fig.~\ref{fig:resolution_model}). Alternatively, an additional memory footprint decrease of $36\%$ can be reached while preserving an accuracy of $90\%$. 

The measurement results of FlexSpIM are presented in Fig.~\ref{fig:results}. Nominal measurement conditions correspond to a \mbox{1.1-V} core voltage, a \mbox{157-MHz} system clock that defines complete CIM operations, and a \mbox{942-MHz} internal clock that splits internal CIM phases, as illustrated in Fig. \ref{fig:combined_cim_details}(c). First, when operands are mapped over all available columns using a single-row shape and an increasing resolution equal for both weights and membrane potentials, the energy per operation increases linearly with the operands resolution. A marginal overhead under $5\%$ appears with an increased resolution due to the carry propagation in the PC adder tree. Next, changing the operand mapping ($N_{R}\times N_{C}$, where $N_{R}$ and $N_{C}$ respectively stand for the number of rows and columns required to store the operand) for a fixed target resolution (e.g., 16 bits) and number of output channels (e.g., 32) modifies the number of CIM columns involved in the computation, as operations spread out sequentially with the number of rows. Compared to a row-wise kernel stacking in \cite{IMPULSE, ISSCC24, NeuroCIM, SpikeCIM, DRAM_CIM, ds_cim, ISCAS22, SNNIM, SRIF}, FlexSpIM’s dedicated operand shaping combined with the PC standby mode for unused columns allows saving up to $4.3\times$ energy per operation, keeping the energy variation across all considered FlexSpIM shapes below $24\%$ (Fig. \ref{fig:results}(a)). This homogeneity mainly originates from the PC standby mode that decreases the energy of inactive columns by $87\%$. A detailed comparison between FlexSpIM and state-of-the-art SNN accelerators is provided in Table \ref{comparison_table}, highlighting $2\times$ better 1-bit-normalized energy efficiency compared to past digital CIM while supporting bitwise granularity on resolution reconfiguration.

\subsection{System Level}
At the system level, accurate evaluation can be obtained by considering a many-macro CIM array architecture with a global on-chip buffer and an external DRAM (Fig. \ref{fig:results}(b)). By extracting the energy efficiency of the complete system, accounting for macro-level measurements, the impact of FlexSpIM’s system-level flexibility is assessed on the six-layer SCNN workload. First, considering the optimum-resolution mappings of FlexSpIM and \cite{ISSCC24} (Fig. \ref{fig:combined_flexible_resolution}(a)), a FlexSpIM-based system composed of 16 CIM macros with HS maximizing the amount of stationary operands achieves an $87-90\%$ energy efficiency gain in the $85-99\%$ input sparsity range. Second, considering the fixed 6-bit weight and 11-bit membrane potential resolutions of IMPULSE \cite{IMPULSE}, a FlexSpIM-based system with 18 CIM macros achieves a $79-86\%$ energy efficiency gain in the $85-99\%$ sparsity range.

\section{Conclusion}
With FlexSpIM, we introduced arbitrary resolution and operand shaping, together with a unified weight/membrane potential memory, to enable CIM-based SNN hardware with the highest resolution and dataflow configurability. With a prototype fabricated in 40-nm CMOS, we demonstrated based on silicon measurements that this flexibility, enabling a $79-90\%$ reduction of energy per operation at the system level, is achieved while maintaining a competitive macro-level trade-off between peak throughput and energy per operation, especially compared to past digital CIM solutions. This work thus underlines the importance of macro-level flexibility to enable significant system-level gains in scalable architectures for real-world edge-vision tasks. 

\section*{Acknowledgment}
This project was co-funded by Prophesee and by the Dutch government as an HTSM-TKI project. The authors would like to thank Douwe den Blanken and Martin Lefebvre, from tapeout support to fruitful discussions.

\end{document}